\documentclass[12pt,apjl]{emulateapj}

\begin{document}


\title{Evidence for accretion in a nearby, young brown dwarf}


\author{Ansgar Reiners\altaffilmark{*}}
\affil{Institut f\"ur Astrophysik, Georg-August-Universit\"at, Friedrich-Hund-Platz 1, D-37077 G\"ottingen}
\email{Ansgar.Reiners@phys.uni-goettingen.de}


\altaffiltext{*}{Emmy Noether Fellow}


\begin{abstract}
  We report on the discovery of the young, nearby, brown dwarf
  2MASS~J0041353$-$562112. The object has a spectral type of M7.5, it
  shows Li absorption and signatures of accretion, which implies that
  it still has a disk and suggests an age below 10\,Myr.  The space
  motion vector and position on the sky indicate that the brown dwarf
  is probably a member of the $\sim$20\,Myr old Tuc-Hor association,
  or that it may be an ejected member of the $\sim$12\,Myr old
  $\beta$\,Pic association, both would imply that
  2MASS~J0041353$-$562112 may in fact be older than 10\,Myr. No
  accreting star or brown dwarf was previously known in these
  associations.  Assuming an age of 10\,Myr, the brown dwarf has a
  mass of about 30\,M$_{\rm Jup}$ and is located at 35\,pc distance.
  The newly discovered object is the closest accreting brown dwarf
  known.  Its membership to an association older than 10\,Myr implies
  that either disks in brown dwarfs can survive as long as in more
  massive stars, perhaps even longer, or that star formation in
  Tuc-Hor or $\beta$~Pic occured more recently than previously
  thought.  The history and evolution of this object can provide new
  fundamental insight into the formation process of stars, brown
  dwarfs, and planets.
\end{abstract}





\section{Introduction}

At the age of $\sim$1\,Gyr, a typical brown dwarf is about 10$^4$
times less luminous than the Sun \citep{Burrows97}, with the
consequence that detailed investigation of old brown dwarfs is only
possible within a distance of a few pc. At young ages, brown dwarfs
are warmer and larger so that they can be seen in star forming regions
as distant as 100\,pc or more. Only a few young brown dwarfs are known
at distances closer than 100\,pc.  In the solar neighborhood, a number
of stars younger than a few 100 Myr are known \citep{Wichmann03,
  McGehee08}, the majority belonging to young associations.  Only rare
cases exist of stars that are apparently born in isolation, these are
usually high mass stars \citep[e.g.,][]{Aarnio08}.

Two competing formation scenarios exist for brown dwarfs, one
suggesting that brown stars form the same way stars do
\citep{Padoan04}, another suggesting that brown dwarfs are stellar
embryos ejected from the collapsing cloud before they could build up
enough mass to become a star \citep{Reipurth01, Bate03}.  Young stars
and brown dwarfs can harbor accretion disks in which planets can form
\citep{Millan-Gabet07}, so the first few million years of star
formation are particularly interesting for multiple reasons. For
example, the lifetime of an accretion disk puts severe constraints on
the brown dwarf formation scenario and sets the timescale for planet
formation.

The lifetime of accretion is on the order of 10\,Myr, it seems to be
similar for brown dwarfs and stars. For stars, \citet{Mamajek02} found
that only one out of 110 stars showed accretion in the 10--20\,Myr
Lower-Centaurus-Crux subgroup of the nearest region of very recent
star formation, Sco-Cen. In the associations $\beta$ Pic
($\sim$12\,Myr) and Tuc-Hor ($\sim$20\,Myr), no accreting star was
found \citep{Jayawardhana06} implying that the accretion phase in
stars (earlier than spectral type M) probably is shorter than
10--20\,Myr. 

Disk accretion is also observed in brown dwarfs that are members of
very young star forming regions ($<$5--10\,Myr). It was studied in
regions as far away as 150~pc, e.g., $\rho$~Ophiuchus, Upper Scorpius,
Taurus and others \citep{Natta04, Mohanty05, Jayawardhana06,
  Herczeg09}. Only a handful of young objects are known with spectral
types later than M5 (cooler than $\sim$3000\,K) that are at the end of
the disk accretion phase. They are all members of young associations;
three of them are members of TW~Hya ($\sim$8\,Myr): TWA~5B, TWA~26,
and TWA~27.  HR~7329B \citep[$\eta$~Tel\,B,][]{Lowrance00} is a brown
dwarf companion to an A0 star and associated to $\beta$~Pic
\citep{Zuckerman04}. Another young brown dwarf, RX J1857.5-3732W, was
found in the Ext~R~CrA region \citep{Neuhaeuser00}.  Only one of the
five objects, TWA~26, is known to be an accretor \citep{Mohanty05}.
Furthermore, TWA~26 harbors a planet \citep{Chauvin04} and shows
signatures of an outflow \citep{Whelan07}.  This object became a
benchmark in the studies of young brown dwarfs, but at a distance of
$\sim$70pc, the apparent magnitude of TWA~26 is $J \approx 13$, which
makes detailed investigations of the object very difficult. The newly
discovered accreting brown dwarf 2MASS J0041353-562112 (hereafter
2M0041) seems to be very similar to TWA~26 and probably somewhat
older.  At about half the distance, it appears one magnitude brighter.

\section{Data}

In an effort to characterize space motion and stellar activity in
nearby stars \citep[][]{Reiners09}, we observed a sample of ultracool
M7 -- M9 dwarfs with UVES at the Very Large Telescope
\citep{Dekker00}, using a setup centered at 830\,nm that provides a
spectrum in the range 640--1020\,nm in the red arm. UVES was used in
dichroic mode that also provided a spectrum at 376--500\,nm in the
blue arm. Data reduction followed standard procedures implemented in
the ESO CPL pipeline version 4.3.0, this includes bias subtraction, 2D
flat-fielding, and wavelength calibration using ThAr frames.

\section{Li absorption}

Young brown dwarfs of several ten M$_{\rm Jup}$ have spectral types
late-M \citep{Burrows97} and are difficult to distinguish from older
stars of the same spectral type. One method to identify a young brown
dwarf is the detection of a Li absorption line. We detect a strong
absorption line of Li at 6708\,\AA\ in our spectrum of 2M0041
\citep[shown in][]{Reiners09}, which means that it is a brown dwarf
younger than $\sim$200\,Myr, and that it is less massive than
65\,M$_{\rm Jup}$ \citep{Basri00}.

\section{Accretion}

\subsection{Indicators of accretion}

The most important indicator of accretion is a strong and asymmetric
H$\alpha$ emission line. Stars with equivalent widths larger than
10\,\AA\ are often categorized as accreting classical T~Tauri stars
\citep[e.g.,][and references therein]{Jayawardhana03, Muzerolle03},
but this cutoff is higher for low-mass objects because chromospheric
H$\alpha$ is more easily visible because of the decreased photospheric
continuum. A more robust indicator of accretion is profile shape,
objects with broad \emph{and} asymmetric H$\alpha$ lines are likely to
be accretors \citep[e.g.,][]{Muzerolle03}. \citet{Jayawardhana03}
argue that a 10\,\% H$\alpha$ full width of $\ga 200$\,km\,s$^{-1}$ is
a good accretion cutoff. \citet{Natta04} have shown that the H$\alpha$
10\,\% width correlates very well with mass accretion rates derived by
other means. Following \citet{Jayawardhana03} and \citet{Mohanty03},
accretion can conservatively be identified based on asymmetric and
broad H$\alpha$ \emph{and} the detection of other emission lines.

Emission lines of other elements indicative of accretion are lines of
\ion{He}{1}, \ion{O}{1}, and \ion{Ca}{2} \citep[e.g.,][]{Muzerolle98},
most prominent are the lines of \ion{He}{1} $\lambda\lambda$~5876 and
6678, \ion{O}{1} $\lambda\lambda$~7773 and 8446, and lines of the
infrared Ca triplet, Ca~II $\lambda\lambda$~8498, 8542, and 8662. All
lines appear stronger with higher mass accretion rate, \ion{O}{1}
lines are only found in CTTS with very strong accretion; \emph{\.M}
$\ga 10^{-8}$\,M$_\odot$\,yr$^{-1}$ \citep{Muzerolle98}. A
comprehensive discussion of the appearance of emission lines in young
brown dwarfs is given in \citet{Mohanty05}.  They conclude that
emission in \ion{O}{1}, \ion{He}{1}, and \ion{Ca}{2} 8662 seems
reasonably well correlated with accretion in very-low mass stars and
brown dwarfs, and that these lines, while not detected in all low-mass
accretors, appear preferentially associated with accretion, and not
activity.

\citet{Fuhrmeister08} show a spectrum of the mid-M dwarf CN~Leo
observed during a giant flare. They show that all lines observed in
accretors can also be found in non-accretors during a very strong
flare. However, these lines appear very narrow so that even in the
unlikely event of a giant flare, the shape of the H$\alpha$ line is a
good discriminator between activity and accretion, which also implies
that an asymmetric H$\alpha$ line is not likely to be explained by
activity.

Our UVES spectrum of 2M0041 covers H$\alpha$--H$\delta$, \ion{He}{1}
6678, the two \ion{O}{1} lines, the \ion{Ca}{2} infrared triplet, and
Ca~H\&K.  Emission is observed in H$\alpha$--H$\delta$, in the He~I
line, in all three lines of the Ca~II triplet and in Ca~H\&K. The only
lines that do not exhibit detectable emission are the two O~I lines.

\begin{figure}
  \plotone{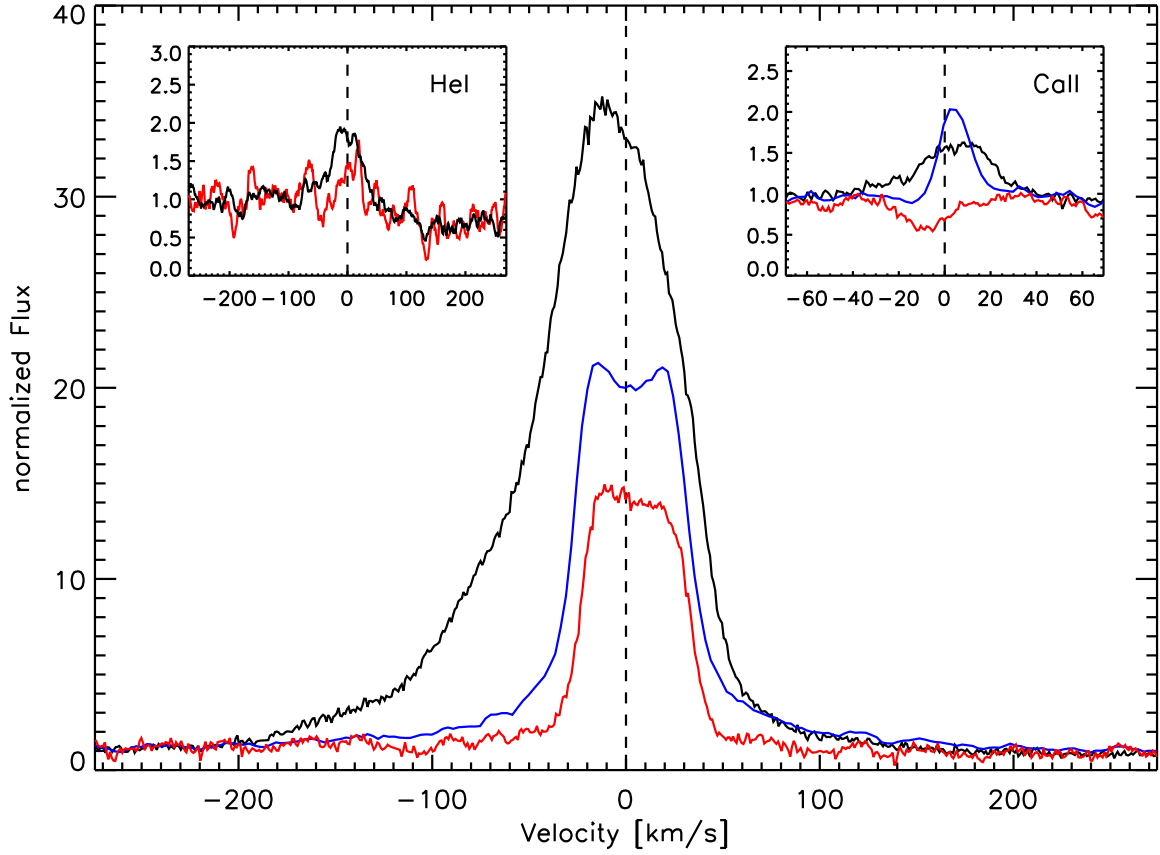}
  \caption{\label{fig:Halpha}Line profiles of H$\alpha$ (main frame),
    He~I 6678 (upper left inset), and Ca~II 8662 (upper right inset).
    The black line shows the spectrum of 2M0041, which is compared to
    profiles of two active low-mass stars, 2MASS~J0752239$+$161215
    (M7.0, red), and 2MASS~J2331217$-$274949 (M8.5, blue). No data at
    6678\,\AA\ is available for the latter star.}
\end{figure}

We show the H$\alpha$ line together with \ion{He}{1} 6678 and
\ion{Ca}{2} 8662 in Fig.\,\ref{fig:Halpha}. For comparison, we
overplot the spectra of two very active late-M dwarfs. The equivalent
width of the H$\alpha$ line of 2M0041 is $-92.7$\,\AA, the line shows
a strongly asymmetric shape.  Both, strength and shape of the
H$\alpha$ line are difficult to explain by chromospheric activity.

The two comparison stars also exhibit strong H$\alpha$ emission lines,
but in both cases, the line is symmetric and relatively narrow.  In
one case, there is even emission observed in \ion{Ca}{2}, but it is
fairly narrow. In the two comparison stars, emission lines are
probably related to strong activity, but even in the case with
detected \ion{Ca}{2} emission, H$\alpha$ emission does not appear
asymmetric or untypically broad for activity. This supports that broad
H$\alpha$ emission lines are not observed in active low-mass stars but
in accretors \citep[see also][]{Muzerolle00}. We conclude that the
broad and strong H$\alpha$ emission in 2M0041 is unlikely to be due to
activity.  Together with the detection of the Li absorption line and
many other emission lines, the most probable explanation is that
2M0041 is accreting.

\subsection{H$\alpha$ variability}

H$\alpha$ was measured before by \citet[][]{PhanBao06}, who found an
equivalent width of $-37.1$\,\AA, and \citet[][]{Schmidt07} report an
equivalent width of $-24.0$\,\AA.  Both values are lower than the new
measurement, which may indicate that during our UVES observation,
2M0041 was observed in a phase of transient accretion. Both other
values are derived from low-resolution spectroscopy, which may affect
the equivalent width measured in such a broad line, especially the
wings are probably difficult to detect at low resolution. This effect,
however, can certainly not account for the large differences between
individual H$\alpha$ observations.

H$\alpha$ variability in accreting low mass stars was investigated by
\citet{Scholz06}, they report H$\alpha$ equivalent widths between
$-13$ and $-387$\,\AA\ for accretors of spectral types M7.5--M8.  The
two measurements of low H$\alpha$ equivalent widths for 2M0041, $-24$
and $-37$\,\AA, are on the lower end of this range, but still not
unusual for similar low-mass accretors \citep[e.g., TWA~26, EqW =
$-27.7$\,\AA,][]{Mohanty03, Mohanty05}.

\subsection{Accretion Rate}

Following \citet{Natta04} and \citet{Mohanty05}, we derive mass
accretion rates for 2M0041 from the H$\alpha$ 10\% width and from the
equivalent width of the \ion{Ca}{2} line at 8662\,\AA. Equivalent
widths of H$\alpha$ and \ion{Ca}{2} 8662, and H$\alpha$ 10\%-width are
summarized together with the other parameters of 2M0041 in
Table\,\ref{tab:2M0041}. We calculate a mass accretion rate of
\emph{\.M} = 10$^{-10.9}$\,M$_\odot$\,yr$^{-1}$ from the H$\alpha$
10\%-width, and 10$^{-10.5}$\,M$_\odot$\,yr$^{-1}$ from the
\ion{Ca}{2} 8662 equivalent width. The rates calculated from different
diagnostics are consistent with each other, and they are typical for a
$\sim$30\,M$_{\rm Jup}$ accreting brown dwarf at an age of a few Myr
\citep{Mohanty05, Herczeg09}.

\section{Distance and space motion}

\begin{figure}
  \plotone{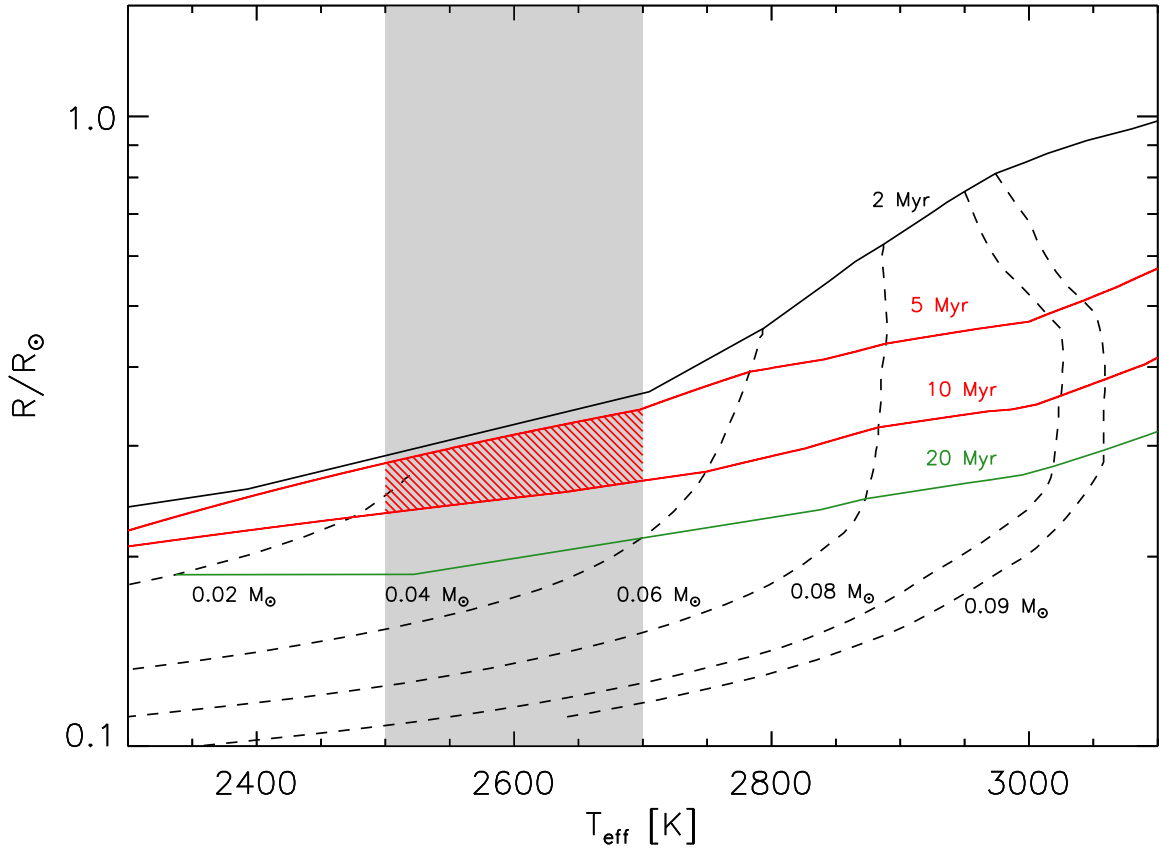}
  \caption{\label{fig:RTeff}Radii of young low-mass stars and brown
    dwarfs as a function of effective temperature. Models are from
    \citet{Baraffe98} and \citet{Baraffe02}. Dashed lines indicated
    the evolution of objects of a certain mass, solid lines are lines
    of same age. The grey area shows the the temperature of M7.5
    objects, the red hatched area shows the radius range for objects
    at this temperature and an age between 5 and 10 Myr. }
\end{figure}

The discovery of accretion in 2M0041 indicates that this object is
very young, probably in the range 5--10\,Myr \citep{Mohanty05}. In
order to test membership to known young associations on the basis of
space motion, we require the distance to the young brown dwarf
together with its proper motion and radial velocity. Radial velocity
was calculated from our spectrum.  Unfortunately, no trigonometric
parallax is available, but proper motion is known \citep{PhanBao06}
and a spectrophotometric distance of 17\,pc is reported under the
assumption that the object is on the Main Sequence \citep{Faherty09}.
At the age of only a few Myr, the radius of 2M0041 is larger than the
radius of a star with the same temperature on the Main Sequence, which
means that it must be at greater distance to appear at the observed
apparent brightness.  We show evolutionary tracks of young brown
dwarfs in Fig.\,\ref{fig:RTeff}. The spectral type of 2M0041 is M7.5
with an uncertainty of $\pm0.5$ \citep{PhanBao06}, i.e., the effective
temperature is $T_{\rm eff} \sim$ 2600\,K\citep[][]{Golimowski04} with
an uncertainty of $\sim 100$\,K.  According to evolutionary models
\citep{Baraffe98, Baraffe02}, 5--10\,Myr old objects at this
temperature are brown dwarfs with masses in the range 20--40\,M$_{\rm
  Jup}$. Their radii are 2--3 times larger than radii of old, Main
Sequence stars with masses of about 90\,M$_{\rm Jup}$ in the same
temperature range. Note that the radius-temperature relation is rather
shallow at this age so that the uncertainty in temperature is
uncritical. If the real age of 2M0041 is between 5 and 10\,Myr, the
distance is roughly 35--50\,pc.

The space motion vector of 2M0041 sensitively depends on the distance
to the object, which is a function of age. We calculated the space
motion of 2M0041 from the proper motion \citep{PhanBao01} and the
radial velocity following \citet{Johnson87}.  We employ the IDL
procedure {\ttfamily gal\_uvw}\footnote{\ttfamily
  http://idlastro.gsfc.nasa.gov/contents.html}, but we use a
right-handed coordinate system with $U$ towards the Galactic center.
Space velocities and distances for ages of 5 and 10\,Myr, and for the
Main Sequence are given in Table\,\ref{tab:2M0041}.

\section{Membership to a young association}

Very young objects only a few Myr old can often be identified as
members of star forming regions and young associations
\citep{Zuckerman04, Torres08}, but only a few associations are as
young as 5--10\,Myr and closer than 50pc. A reliable indicator of
membership to an association is the space motion of an object. Stars
born together usually move into the same direction with only a few
km\,s$^{-1}$ velocity dispersion. Furthermore, young associations can
be concentrated in a narrow region on the sky because they are still
spatially connected and relatively far away. In these cases, the
position of an object on the sky is another important membership test.

Three associations younger than 30\,Myr are significantly closer than
100 pc \citep[we use the values from][]{Fernandez08}, they are TW~Hya
(age 8Myr, $d\sim$60pc), $\beta$~Pic (12 Myr, $\sim$35pc), and Tuc-Hor
(20 Myr, $\sim$50pc).  The distance of 2M0041 at the ages of TW~Hya,
$\beta$~Pic, and Tuc-Hor are approximately 40\,pc, 35\,pc, and 30\,pc,
respectively.  For $\beta$~Pic, the distance to the center of the
association exactly matches the distance to 2M0041 at the age of the
association. For TW~Hya and Tuc-Hor, the values are somewhat
different, but they are not significantly outside the dispersion of
the known association members (and when taking into account the
uncertainties of the spectrophotometric distance).

\begin{figure}
  \plotone{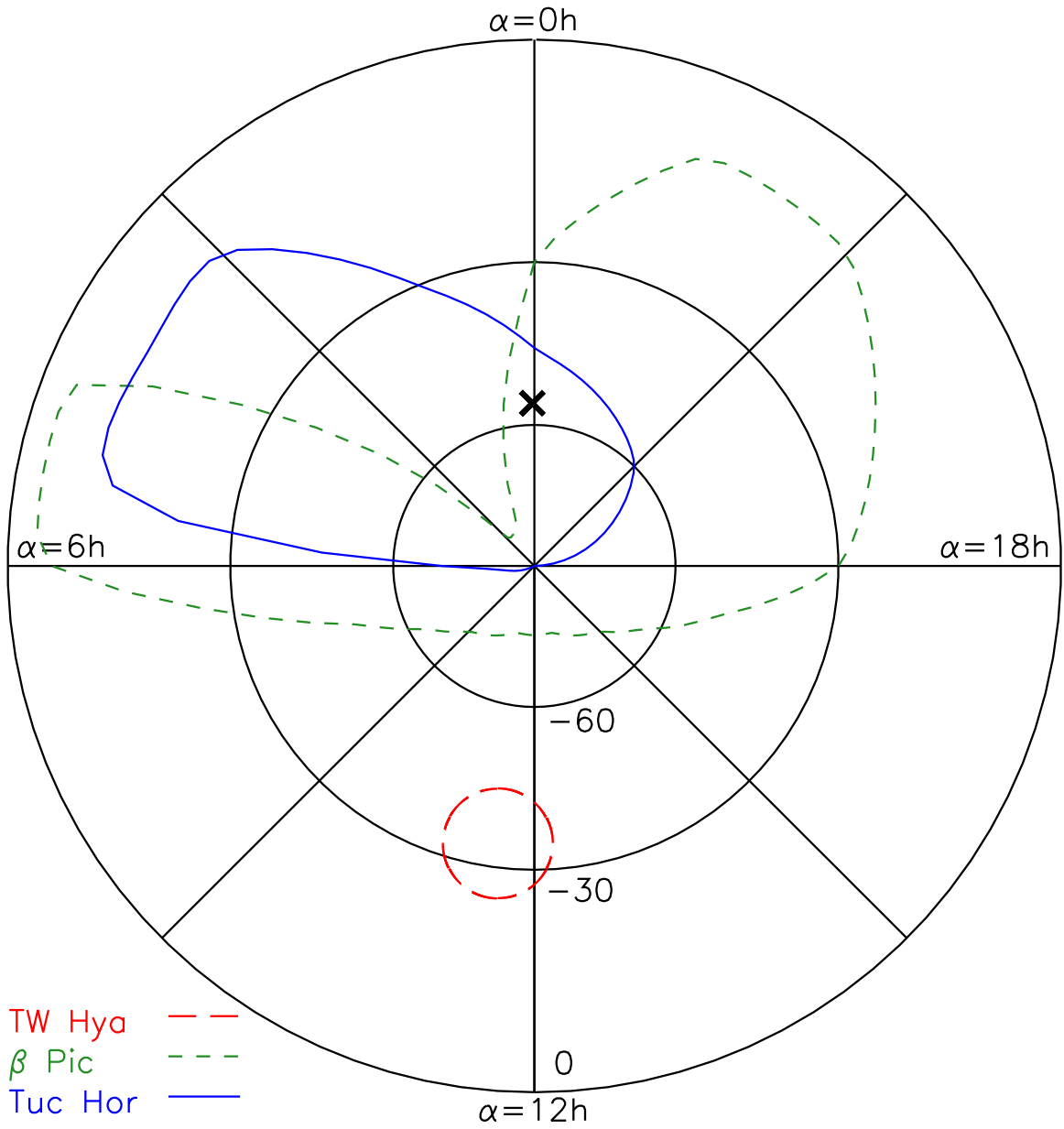}
  \caption{\label{fig:RADec}Celestial polar projection of the Southern
    hemisphere, the position of 2M0041 is indicated with a cross.
    Ellipses show approximate distributions of three young
    associations \citep{Torres08}; red long dashed lines: TW Hya;
    green short dashed lines: $\beta$ Pic; blue solid lines: Tuc Hor.}
\end{figure}

In Fig.\,\ref{fig:RADec}, we show the celestial polar projection of
the Southern hemisphere and the position of 2M0041. We overplot the
loci of the three nearby young associations \citep{Torres08}. Both,
$\beta$ Pic and Tuc-Hor, are distributed over a very large area on the
sky, and both regions cover the position of 2M0041. On the other hand,
TW~Hya is concentrated in a very narrow region far away from the
position of 2M0041. Thus, on the basis of its coordinates, membership
of 2M0041 to TW~Hya can be ruled out.

\begin{figure*}
  \center
  \mbox{
    \resizebox{.5\hsize}{!}{\includegraphics[]{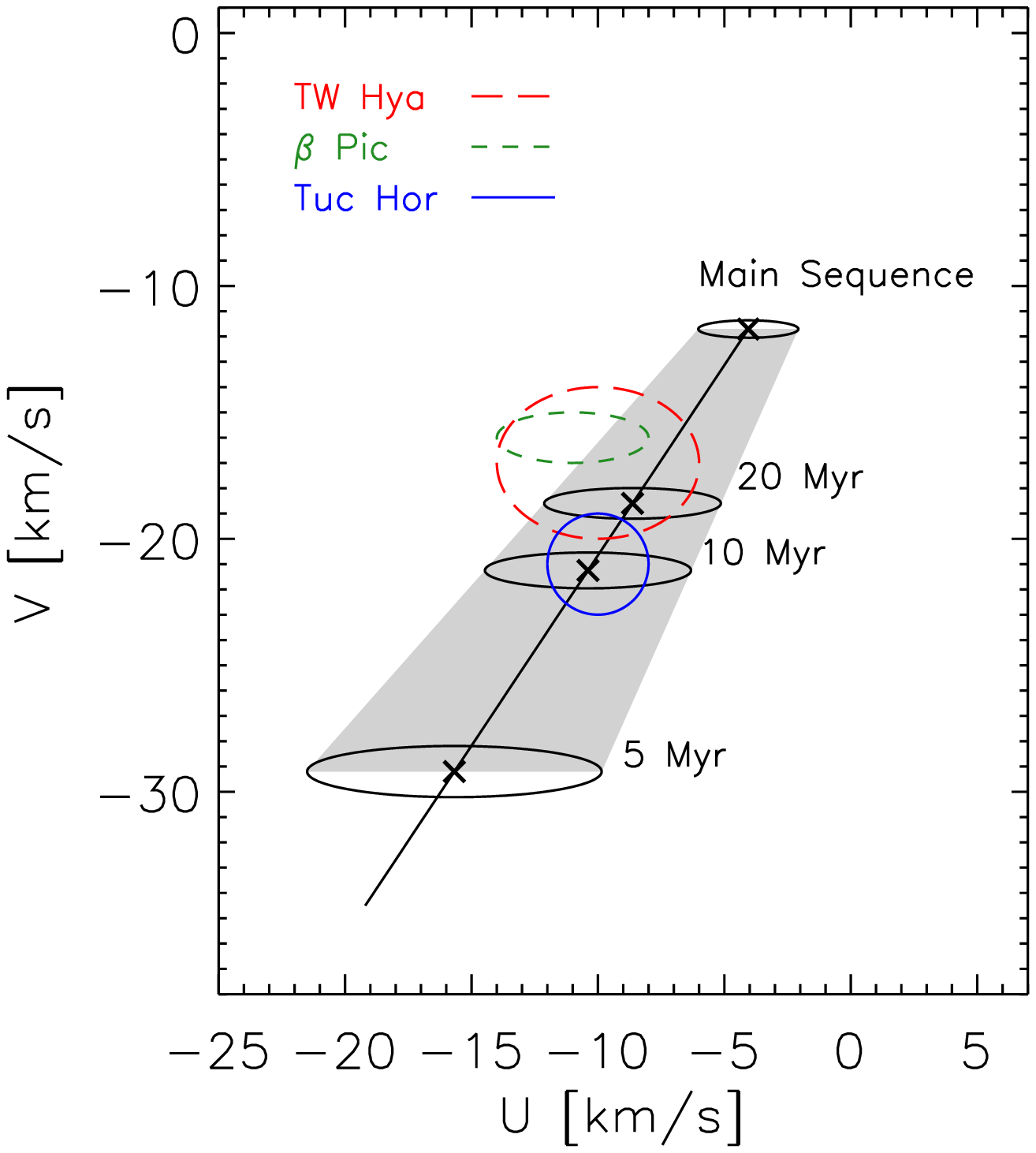}}\quad
    \resizebox{.475\hsize}{!}{\includegraphics[]{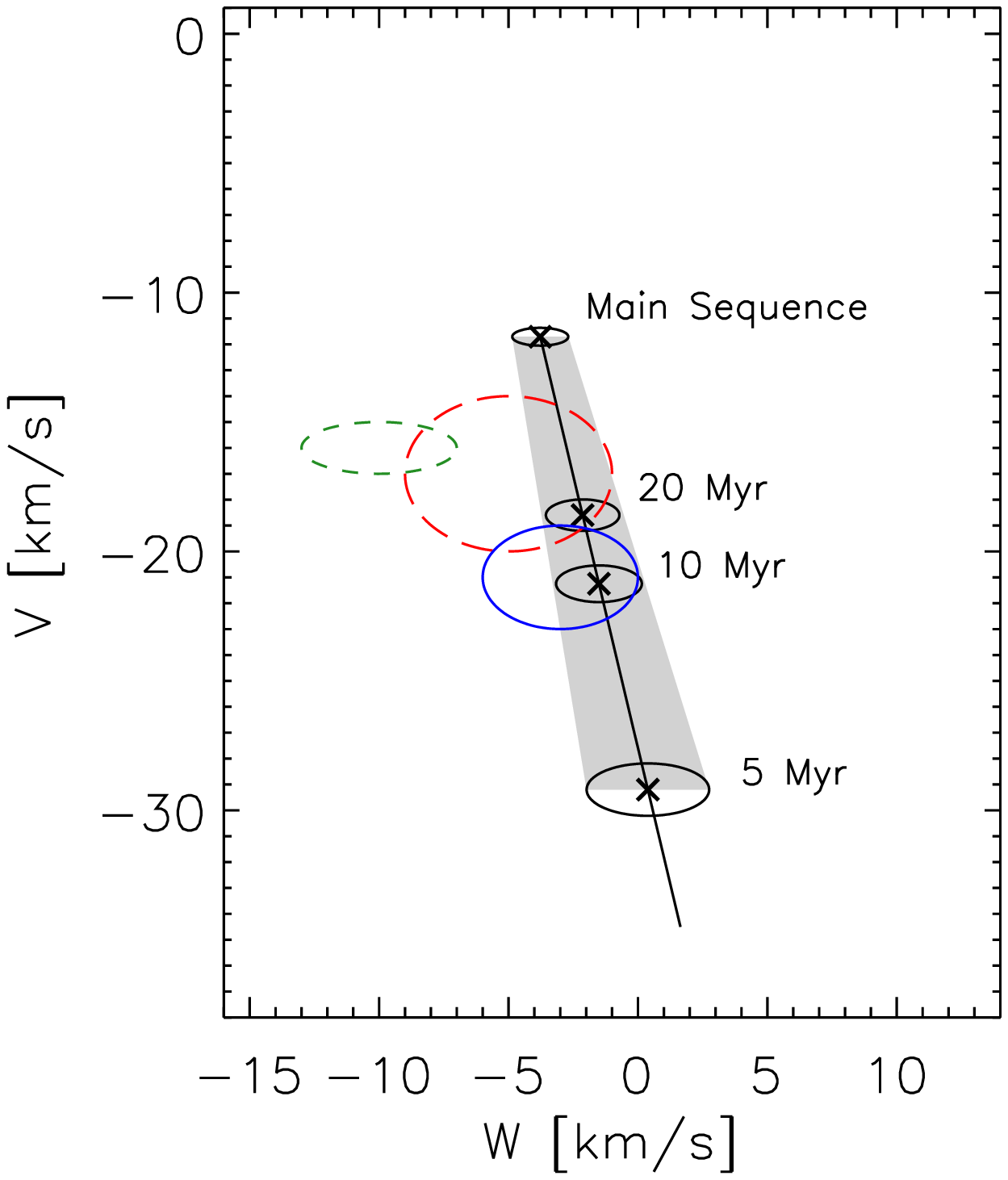}}
  }
  \caption{\label{fig:UVW}Heliocentric space velocities $U, V, W$
    plotted for different ages of 2M0041.  Ellipses show uncertainties
    from uncertainties in proper motion.  The approximate position of
    three young associations are shown as coloured ellipses
    \citep{Fernandez08}; red long dashed lines: TW Hya; green short
    dashed lines: $\beta$ Pic; blue solid lines: Tuc Hor.}
\end{figure*}

The space motion of 2M0041 for different ages is plotted together with
the velocities of TW~Hya, $\beta$~Pic, and Tuc-Hor in
Fig.\,\ref{fig:UVW}. The age-dependent values of $U$, $V$, and $W$
follow straight lines in both diagrams. The space motion vector of
$\beta$~Pic is about 10\,km\,s$^{-1}$ away from the most likely
position of 2M0041, assuming an age of 10 Myr. Thus, from $V$ and $W$
velocities, membership of 2M0041 to $\beta$~Pic is rather unlikely. It
may, however, have been ejected from the main association so that it
moves into a slightly different direction.

Assuming an age of 10--15\,Myr, the space motion of 2M0041 very well
matches the space motion of Tuc-Hor in all three velocities. Together
with the signatures of youth and its position on the sky, we conclude
that 2M0041 is very likely a member of the Tuc-Hor association, that
its age is probably in the range 10--15\,Myr, and that the distance to
the object is around 35\,pc.

\section{Summary}

We found evidence for accretion in a nearby brown dwarf that is
probably a member of the Tuc-Hor association. The most probably age of
Tuc-Hor is around 20\,Myr, but ages in the range 10--40\,Myr are
reported in the literature \citep[see][]{Fernandez08}. If 2M0041 was
formed together with Tuc-Hor (or even $\beta$\,Pic), it is probably at
least 10~Myr old. The detection of accretion implies that disks in
brown dwarfs can survive at least as long as in more massive stars. If
the age is even 15\,Myr or more (which would imply a distance of 30~pc
or less), it means that disks around brown dwarfs may actually live
longer than around stars, which would have serious implications for
brown dwarf formation and for planet formation around brown dwarfs. An
alternative explanation is that the age of 2M0041 is below 10~Myr and
that it formed later than the members of Tuc-Hor or $\beta$~Pic. In
this case, one has to explain how the brown dwarf reached its current
position in the Galaxy, far away from very younger star forming
regions.

In either case, 2M0041 is the closest brown dwarf with evidence for
accretion known so far. It promises important insight into brown dwarf
formation theory and the physics of young low mass objects.


\acknowledgements

Based on observations collected at the European Southern Observatory,
Paranal, Chile, PID 080.D-0140. A.R acknowledges financial support as
an Emmy Noether Fellow of the DFG under DFG RE 1664/4-1.





\begin{deluxetable}{lrrrrr}
  \tablecaption{\label{tab:2M0041}Parameters for 2M0041; the lower
    part shows radius, distance, and space velocities for three
    different ages}
  \tablewidth{0pt} \tablehead{Parameter & \multicolumn{3}{c}{Value}}
  \startdata
  \multicolumn{2}{l}{R.A. (J2000.0)}  & \multicolumn{2}{r}{$00^h41^m35\fs39$}\\
  \multicolumn{2}{l}{Decl. (J2000.0)}  & \multicolumn{2}{r}{$-56\degr21\arcmin12\farcs77$}\\[1mm]
  \multicolumn{2}{l}{$pm_{\rm R.A.}$ \hfill [mas\,yr$^{-1}$]}  & \multicolumn{2}{r}{$121 \pm 19$}\\
  \multicolumn{2}{l}{$pm_{\rm Dec} $ \hfill [mas\,yr$^{-1}$]}  & \multicolumn{2}{r}{$-64 \pm 19$ }\\[1mm]
  \multicolumn{2}{l}{2 MASS $J$ \hfill [mag]} &   \multicolumn{2}{r}{$11.96 \pm 0.02$}\\
  \multicolumn{2}{l}{2 MASS $H$ \hfill [mag]}&   \multicolumn{2}{r}{$11.32 \pm 0.02$}\\
  \multicolumn{2}{l}{2 MASS $K_s$ \hfill [mag]} & \multicolumn{2}{r}{$10.86 \pm 0.03$}\\
  \multicolumn{2}{l}{$v_{\rm{rad}}$ \hfill [km\,s$^{-1}$] } & \multicolumn{2}{r}{$6.8 \pm 1$}\\[1mm]
  \multicolumn{2}{l}{H$\alpha$ EqW \hfill [\AA]} & \multicolumn{2}{r}{$-92.7$}\\
  \multicolumn{2}{l}{H$\alpha$ 10\% width \hfill [km\,s$^{-1}$] } & \multicolumn{2}{r}{200.6}\\
  \multicolumn{2}{l}{log \emph{\.M} (H$\alpha$ 10\%) \hfill [log(M$_{\odot}$\,yr$^{-1}$)]} & \multicolumn{2}{r}{$-10.9$}\\[1mm]
  \multicolumn{2}{l}{\ion{Ca}{2} 8662 EqW \hfill [\AA]} & \multicolumn{2}{r}{$-0.8$}\\
  \multicolumn{2}{l}{log Ca flux\tablenotemark{a} \hfill [log(ergs\,s$^{-1}$\,cm$^{-2}$)]}& \multicolumn{2}{r}{$4.83$}\\
  \multicolumn{2}{l}{log \emph{\.M} (Ca) \hfill [log(M$_{\odot}$\,yr$^{-1}$)]} & \multicolumn{2}{r}{$-10.5$}\\[1mm]
  \multicolumn{2}{l}{\ion{He}{1} 6678 EqW \hfill [\AA]} & \multicolumn{2}{r}{$-1.1$}\\
  \multicolumn{2}{l}{\ion{Ca}{2} 8498 EqW \hfill [\AA]} & \multicolumn{2}{r}{$-1.0$}\\
  \multicolumn{2}{l}{\ion{Ca}{2} 8542 EqW \hfill [\AA]} & \multicolumn{2}{r}{$-2.0$}\\
  \hline
  \noalign{\smallskip}
  Model Age\hfill[Myr] & 5 & 10 & MS\\
  $R$\hfill[R$_{\odot}$] & 0.34 &  0.23 & 0.11\\
  d\hfill[pc]  & 50 & 35 & 17\\
  $U$\hfill[km\,s$^{-1}$] & $-15.7 \pm 5.8$ & $-10.4 \pm 4.1$ & $- 4.1 \pm 2.0$\\
  $V$\hfill[km\,s$^{-1}$] & $-29.2 \pm 1.0$ & $-21.2 \pm 0.7$ & $-11.7 \pm 0.3$\\
  $W$\hfill[km\,s$^{-1}$] & $  0.4 \pm 2.4$ & $ -1.5 \pm 1.7$ & $ -3.8 \pm 0.8$
  \enddata
  \tablenotetext{a}{flux at stellar surface}
\end{deluxetable}

\end{document}